\documentclass[a4paper,fleqn,usenatbib]{mnras}
\newcommand{\jb}{$J$-band}
\newcommand{\msun}{M$_{\odot}$}

\usepackage{txfonts}

\usepackage[T1]{fontenc}
\usepackage{ae,aecompl}

\usepackage{graphicx}	
\usepackage{amssymb}	

\title[Are there GC-like abundance patterns in YMCs?]{Searching for GC-like abundance patterns in young massive clusters\thanks{Based on observations made with ESO telescopes at the La Silla Paranal Observatory under programme ID 084.B-0468(A).}}
\author[I. Cabrera-Ziri et al.]{I. Cabrera-Ziri$^{1,2}$\thanks{ICZ: icabrera@eso.org},   C. Lardo$^2$, B. Davies$^{2}$, N. Bastian$^{2}$, G. Beccari$^1$, S. S. Larsen$^3$,\linebreak \newauthor S. Hernandez$^3$
\\
$^{1}$ European Southern Observatory, Karl-Schwarzschild-Stra{\ss}e 2, D-85748 Garching bei M{\"u}nchen, Germany\\
$^{2}$ Astrophysics Research Institute, Liverpool John Moores University, 146 Brownlow Hill, Liverpool L3 5RF, UK\\
$^{3}$ Department of Astrophysics / IMAPP, Radboud University, PO Box 9010, NL-6500 GL Nijmegen, the Netherlands
}

\date{Accepted XXX. Received YYY; in original form ZZZ}

\pubyear{2016}

\begin{document}
\label{firstpage}
\pagerange{\pageref{firstpage}--\pageref{lastpage}}
\maketitle

\begin{abstract}
Studies during the last decade have revealed that nearly all Globular Clusters (GCs) host multiple populations (MPs) of stars with a distinctive chemical patterns in light elements. No evidence of such MPs has been found so far in lower-mass ($< \sim 10^4$ \msun) open clusters nor in intermediate age (1--2 Gyr) massive ($> 10^5$ \msun) clusters in the Local Group. Young massive clusters (YMCs) have masses and densities similar to those expected of young GCs in the early universe, and their near-infrared (NIR) spectra are dominated by the light of red super giants (RSGs). The spectra of these stars may be used to determine the cluster's abundances, even though the individual stars cannot be spatially resolved from one another.
We carry out a differential analysis between the Al lines of YMC NGC 1705: 1 and field Small Magellanic Cloud RSGs with similar metallicities. We exclude at high confidence extreme [Al/Fe] enhancements similar to those observed
 in GCs like NGC 2808 or NGC 6752. However, smaller variations cannot be excluded.
\end{abstract}


\section{Introduction}
\label{sec:intro}

Today it is clear that nearly all globular clusters (GCs) host multiple stellar populations (MPs)\footnote{With the exception of Ruprecht 106 cf. \cite{Villanova13}.}, as inferred through star-to-star variations in the abundances of some light elements, e.g. the characteristic Na--O anti-correlation (e.g. \citealt{Carretta09}, hereafter C09; \citealt{Roediger14}). Some of these chemical patters are also responsible for the complex colour-magnitude diagram (CMD) of GCs showing multi-modal main sequences, sub-giant branches, etc (c.f. \citealt{Piotto15}, P15 hereafter). The detection of MPs within GCs has consequently challenged our understanding of the very origins of these clusters.


Several scenarios have been put forward to explain the presence of MPs in GCs, with many requiring multiple generations of stars in order to explain the observed discrete sequences in the CMDs and the associated peculiar chemical abundance patterns. The basic hypothesis is that a second generation of stars is born during the early life of the GC from the ejecta of some first generation stars which are polluted in a way that accounts for the signature light elements abundance patterns observed in old GCs today. However, none of the proposed scenarios appear to work. Among the most severe handicaps we find the ``mass-budget problem"
, a collective disagreement between the predicted abundance patterns and the observed ones, and also the lack of ability to reproduce the rather constant value of enriched to non-enriched stars (cf. \citealt{BCZS15, BL15}). Since none of the proposed scenarios appear to be viable, it is important to determine when/where such MPs exist, which may help determine their origin.

None of the models seeking to explain MPs within GCs explicitly invoke ``special" conditions (for example, conditions only found in the early universe), suggesting that the same mechanisms should be operating in young massive clusters (YMCs) today. This makes YMCs ideal places to test GC formation theories (e.g. \citealt{Sollima13}). Additionally, since both metal-rich (bulge) and metal-poor (halo) GCs have been observed to host MPs, and since they likely formed in very different environments and at different redshifts (e.g. \citealt{BS06, Kruijssen14}), it appears likely that the process of the formation of MPs is related to the clusters themselves, and not their host environment (cf. \citealt{Renzini13}). Consequently, the same MPs should be observable in YMCs forming in the present day.

To date, there has been no conclusive evidence of multiple episodes of star formation in YMCs.
That is, no evidence of ongoing star formation has been found in a study of $\sim 130$ young (10--1000 Myr) massive ($10^4 - 10^8$ \msun) clusters \citep{Bastian13}. Nor has any evidence been found of gas reservoirs within YMCs that could fuel extended star-formation episodes with the masses suggested by the formation scenarios listed above (e.g. \citealt{BS14,cz15}). Additionally, no evidence of age spreads have been found either from the analysis of the CMDs or integrated light of YMCs (e.g. \citealt{Niederhofer15, cz14,cz16a,cz16b}). These studies have called into question the proposed scenarios, however, \emph{they have not tested whether the distinctive chemical patterns characterising MPs are present within these clusters.}

So far, no evidence of the characteristic GC abundance patterns has been found in the Milky Way open clusters in the Galactic disk \citep{deSilva09,Pancino10,Magrini14,Magrini15,Maclean15}; more massive ($\sim 10^4$ \msun) old open clusters were targeted by \cite{Bragaglia12} (Berkeley 39, $\sim 6$ Gyr) and \cite{Bragaglia14,Cunha15} (NGC 6791, $\sim 9$ Gyr). None of these studies found signs of stars with ``polluted chemistry". Furthermore, the LMC intermediate-age (1--2 Gyr, $> 10^5$ \msun) clusters NGC 1806, 1651, 1783, 1978, and 2173 do not show signs of GC-like abundance patterns \citep{Mucciarelli:2008p2339,Mucciarelli:2014p2575}. 
 \emph{This lack of evidence of MPs in younger clusters has been suggested to be due to the fact that they have lower mass/density than the ancient GCs.} However, these studies have shown that a sharp mass/density limit does not apply, as there exist overlap of these properties between the samples with and without MPs.

There are clusters that are forming in the nearby universe with masses well in excess of $10^6$ \msun (e.g.\ in the Antennae Galaxies; \citealt{Whitmore10}). These clusters 
 have properties similar to those expected for young GCs (cf. \citealt{SS98,PZ10}), and hence we may expect that they also have formed MPs.

The traditional method to find and quantify the signatures of star-to-star abundance variation has been via high-resolution spectroscopy of individual stars in a cluster. Young clusters with masses and densities similar to early GCs are only found in external galaxies. Due to their distances (tens of Mpc), a detailed abundance analysis of their individual stars is not possible with the instrumentation currently available.
 While we cannot resolve these clusters into their constituent stars, here we will present a method, the \jb\ technique, devised to allow us to look for chemical anomalies, i.e. the MPs, within these clusters using their integrated near-infrared (NIR) spectrum. 

\section{RSG stars and YMCs}
\label{sec:jband}

\cite{Davies10} developed a technique, a.k.a \jb\ technique, whereby chemical abundances of red super giants (RSGs) may be extracted from a narrow spectral window around 1\micron\ from low resolution data ($R\sim3000$). The method is therefore extremely efficient, allowing stars at large distances to be studied, and so has tremendous potential for extragalactic abundance work. Several studies have shown that the \jb\ technique rivals the precision ($\pm0.1$ dex) of metallicity measurements using Blue Super Giants (frequently used to determine metallicities beyond the Local Group) and is applicable over similar distances (several Mpc) with existing instruments \citep{Gazak15,Lardo15}.

The effective temperature of the RSGs are constant to within $\pm200$ K \citep{Davies13} and does not depend on the stellar metallicity \citep{Gazak15}. Within a coeval cluster the RSGs all have similar luminosities and virtually identical masses, and therefore similar gravities.
 \emph{In other words, for a given metallicity and stellar mass, RSGs have almost identical spectra in the \jb.} 

When a YMC reaches an age of $\sim7$ Myr, the most massive stars which have not yet exploded as supernovae will be in the RSG phase. For a cluster with an initial mass of $10^5$ \msun, there may be more than a hundred RSGs present which dominate the cluster's light output in the NIR, contributing 90\% - 95\% of the of the NIR flux \citep{Larsen06,Gazak13}. \emph{As RSGs have all the same effective temperature, the integrated light spectrum of a YMC can be analysed in the same way as a single RSG spectrum (cf. \citealt{Gazak14}).}
As a result, the \jb\ technique can be applied to unresolved star clusters as well as individual stars.

Our method to search for MPs in YMCs exploits the fact that ``RSGs all look the same". So, if the mechanisms responsible of the GCs abundance variations (i.e. MPs) are in force today, and we were to compare RSGs from the field with the integrated light spectra of young clusters dominated by RSGs, we would expect to find both spectra very similar (i.e. to have similar Fe, and most of metals), and only see differences in the abundances of the elements that vary within GCs (i.e. C, N, O, Na and Al). These abundance differences will be due to the contribution of ``polluted" RSGs to the integrated light of the cluster. 

\section{Data}
\label{sec:data}

We will focus this pilot study on a young ($\sim15$ Myr) massive ($\sim10^6$ \msun) cluster in NGC 1705, a blue compact dwarf galaxy 5.1 Mpc away, with a metallicity similar to the Small Magellanic Cloud (SMC, e.g. \citealt{Annibali09}). For our analysis we use archival NIR spectroscopic data of the of cluster NGC 1705: 1 obtained on Nov 23rd 2009 with the XSHOOTER spectrograph on the Very Large Telescope under ESO programme number 084.B-0468(A) (PI S. S. Larsen). The cluster was observed in a single AB nodding cycle, using the $0.9\times\!11\arcsec$ slit placed at parallactic angle. The total exposure time was 600 s, during which the airmass increased from 1.32 to 1.35 and the seeing from $0\farcs98$ to $1\farcs17$. Flux standard stars were also observed and to correct for the atmospheric absorption in the NIR, telluric standard stars of spectral type late-B were observed within one hour of each science target. The data reduction consisted in subtraction of bias and dark frames, flat-fielding, order extraction and rectification, and flux and wavelength calibration, this was carried out using the standard ESO Reflex pipeline version 2.6.0. At the end of this we achieved a SNR at the \jb\ of >100 per spectral bin.

\section{Analysis}
\label{sec:anal}

\begin{figure*}
\includegraphics[width= 180mm,height=42mm]{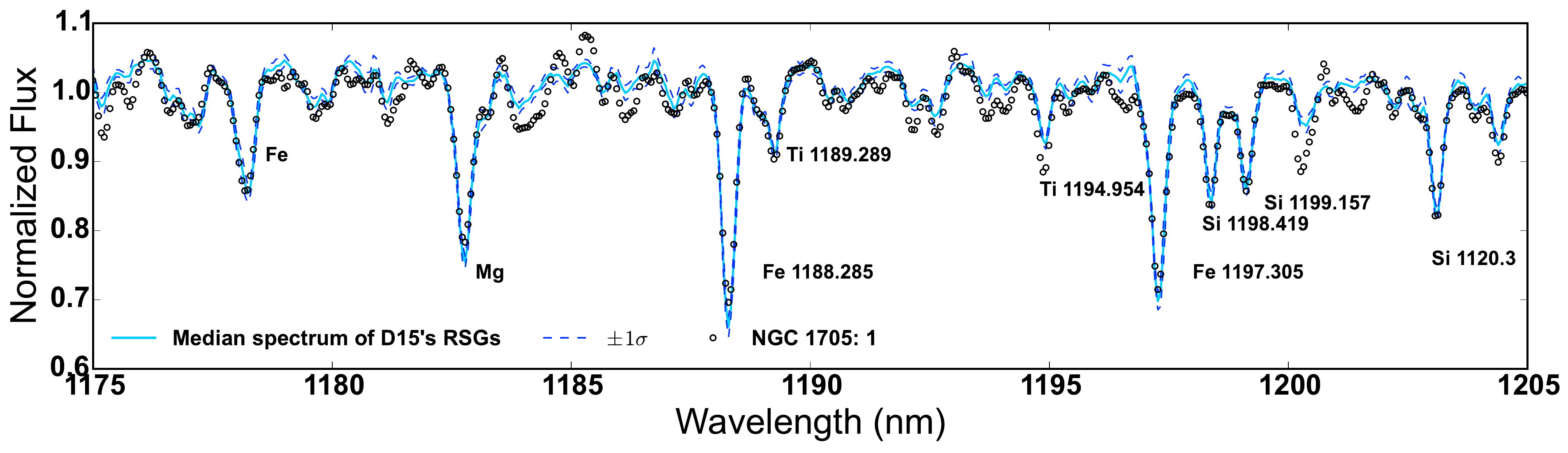}
\caption{In cyan we show (part of) the median \jb\ spectra of the RSGs from D15. The blue dashed lines denote $\pm 1 \sigma$ spectra of the D15 stars. The cluster's spectrum is shown in black circles. All spectra have been downgraded to the same resolution ($R=4000$). As expected, we find a good agreement between the metals lines of our cluster and the median RSG spectrum. Mg remains constant from star-to-star in most GCs, however, [Mg/Fe] it has been observed to be depleted in the stars of a few GCs (e.g. NGC 2808 and 7078, cf. C09). There is no evidence supporting such depletion in this cluster.}
\label{specjband}
\end{figure*}


We performed a differential analysis, comparing the \jb\ spectrum of NGC 1705: 1 and a representative RSG median spectrum with similar metallicity. 
 For this, we first built a suitable comparison sample by selecting SMC RSGs spectra spanning a metallicity range between $-0.24 \le \mbox{[Z]} \le -0.72$, similar to that of NGC 1705. 
These spectra were taken from \cite{Davies15}, hereafter D15. 
Then we calculated a median and standard deviation spectra ($\sigma$) of all the RSGs in our sample. As these RSGs do not belong to dense/massive clusters we do not expect to see in them the chemical patterns characteristic of GCs stars. Hence, if the mechanisms that are responsible for the distinct abundance patterns of GCs are still acting today in the universe, we should observe that most metal lines (i.e. lines of species that are not seen to vary strongly in GCs -- anything other than C, N, O, F, Na, Mg, Al, Si, K)
should be the same on the D15 median RSG spectrum as on NGC 1705: 1, as they show no significant variation on GC stars. On the other hand, chemical species that have large variations from solar-scaled abundances in GCs (e.g.\ Na, O, Al and N) should be enhanced/depleted accordingly in the spectrum of NCG 1705: 1 with respect to D15 median RSG spectrum. For our analysis, all the spectra were homogenized in terms of their spectral resolution.

\begin{figure}
\includegraphics[width= 84mm]{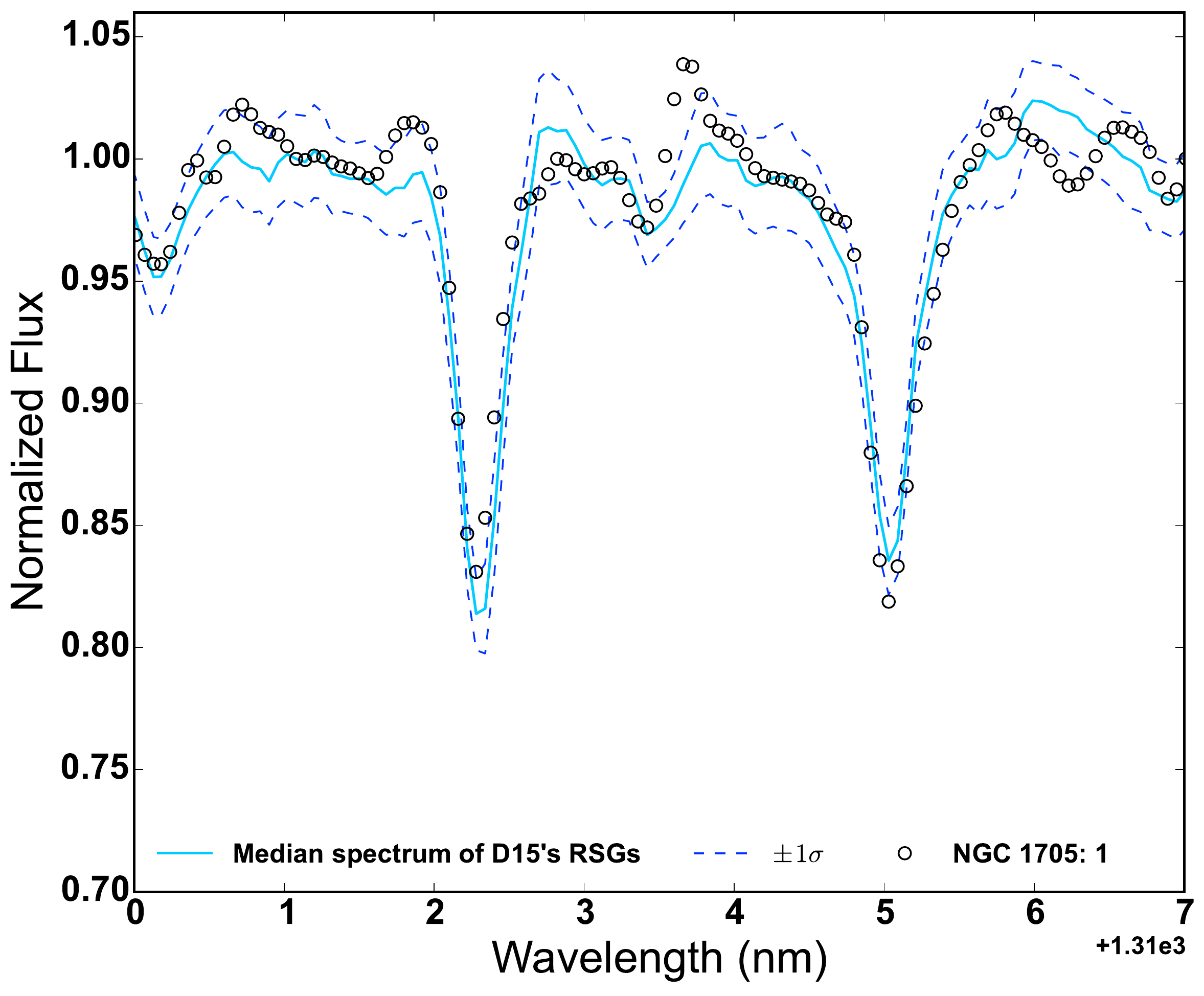}
\caption{Similar to Fig. \ref{specjband} but centred around two prominent Al lines (1312.338 and 1315.071 nm). These lines in NGC 1705: 1 and D15's RSGs are in agreement within the observational uncertainties. This is not what would be expected if the Al of this cluster were enhanced significantly like it is often observed in some GCs.
}
\label{Al:RSG}
\end{figure}


In Fig. \ref{specjband}, we show how the median RSGs spectrum of D15's stars compare to the spectrum of our cluster in the region between 1175--1205 nm. As expected, we find a remarkable agreement between the metal lines of the field RSGs and our cluster, indicating that NGC 1705: 1 has a metallicity\footnote{Assuming that Fe, Mg, Si and Ti abundances from individual lines are representative of the metallicity Z.} similar to RSGs stars in the SMC. We also made use of the \jb\ technique (\citealt{Davies10}, cf. \S\ref{sec:jband}) to make a quantitative assessment of the metallicity of this cluster and its uncertainty, yielding [Z] $=-0.4\pm 0.1$ dex.

\subsection{Expectations from different GCs}
\label{sec:GCmod}

Al is the only element, with strong lines in the NIR spectrum of RSGs, that shows significant abundance variations in GC stars. Na and O are other elements with evidence of strong variations from star-to-star in GCs. However, there are no Na lines in the \jb\ spectra of RSG, and for these stars most of the O is locked up in molecules and so is difficult to measure directly. RSGs in a cluster of this age ($\sim 15$ Myr) still contribute to $\sim50\%$ of the light in the regions where optical Na and O lines (like Na{\sc i} doublets at 5682-88 \AA\ and 6154-60 \AA; and [O{\sc i}] 6300 and 6363 \AA\ lines) are found. However, carrying on a similar analysis as in the \jb\ is not possible, as one would need to consider how the rest of the stellar population (i.e. all other stars that are not RSGs) affect these spectral regions. Additionally, the variations in these particular lines (i.e. difference between a star with ``polluted chemistry'' and a regular one) are not as large as the ones observed in the NIR Al lines. All this makes this kind of analysis significantly more complex to carry out on such features.

Extreme differences (>1 dex) in [Al/Fe] have been observed in the stars of some GCs e.g., M~54, NGC 2808, M~80, NGC~6752 and NGC~6139 (see below). We show in Fig. \ref{Al:RSG}, that the Al lines from NGC 1705: 1 do not show a significant enhancement when they are compared to field RSGs of similar metallicity. This is contrary to what one would expect if there were RSGs with a range of Al abundances similar to that of the GCs mentioned above. To illustrate this point, we have computed exploratory model spectra of RSGs in order to investigate how these differences in the Al lines should look like. For this, we used MARCS model atmospheres \citep{Gustafsson08}, and their spectra were computed with TURBOSPECTRUM \citep{Plez12}. We assumed the following parameters for our models: $T_{\mbox{eff}}=3800$ K, $\log g = 0.5$ dex, [Z] $= -0.5$ dex, $\xi=2.0$ km s$^{-1}$, [$\alpha$/Fe] = 0.0 dex\footnote{While the assumption of a solar-scaled composition is reasonable, we expect that the errors arising from the assumption of a solar-scaled rather than an $\alpha$-enhanced one has a negligible impact (within the quoted 0.1 dex errors in metallicity) on the metallicity determination \citep{Lardo15}.} and vary the Al abundances simulating the enhancement expected of some GCs. We note that these models were not intended to measure absolute Al abundances, rather to estimate what type of variations we might expect if there were GC-like chemical anomalies in this YMC.

\begin{figure}
\includegraphics[width= 84mm]{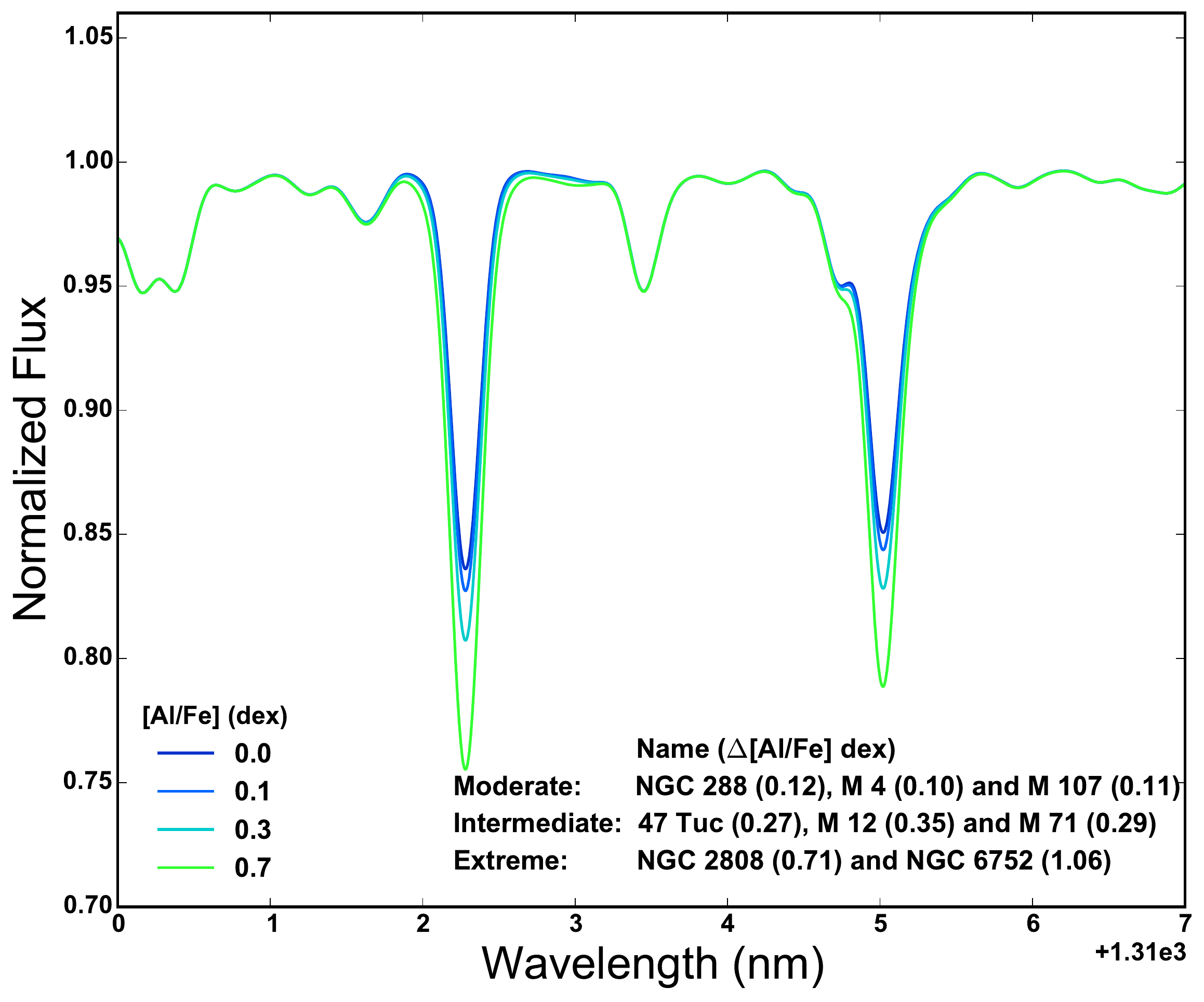}
\caption{RSG models with solar-scaled (0.0 dex -- as expected for the D15 RSGs) and enhanced (>0.0 dex) [Al/Fe] values. We can exclude at high confidence that NGC 1705: 1 have extreme values of [Al/Fe] (like NGC 2808 or NGC 6752) that depart from the solar-scaled [Al/Fe] abundance of the RSGs form D15, cf. Fig. \ref{Al:RSG}.}
\label{Al:models}
\end{figure}


With these models we synthesised how the NIR spectra of GCs with different ranges of [Al/Fe] spread would look like at young ($\sim10$ Myr) ages. We assumed for our calculations that 1) the ratio of enriched to non-enriched stars of young GCs is the same as the one observed today  (i.e. $\sim$70:30\% cf. \citealt{Carretta09a,BL15}) and 2) that all RSGs that dominate the NIR light of the cluster at this age had the same luminosity. 

For this experiment we have compiled a homogenous sample of [Al/Fe] abundances for 25 GCs from the literature. The median number of stars per GC analysed in this sample was 12 (with a minimum of 4 stars for M79 and a maximum 100 stars for NGC 6752). We divide the [Al/Fe] spread observed in GCs in three ranges: moderate (e.g. NGC 288, M~4, M~79, or M~107); intermediate (e.g. 47~Tuc, M~12 or M~71) and extreme (e.g. NGC 2808, NGC 6752 and M~54). We define these ranges according to the difference, $\Delta$[Al/Fe] = mean([Al/Fe])--min([Al/Fe]), between the mean [Al/Fe] abundance and the pristine [Al/Fe] abundance for the stars in these clusters. The moderate, intermediate and extreme ranges have values of $\Delta$[Al/Fe] = 0.1, 0.3 and 0.7 dex respectively. The clusters in this sample, their spreads and references are listed in Table \ref{tab1}.

\begin{table*}
\caption{Metallicity and [Al/Fe] spreads ($\Delta$[Al/Fe], standard deviation and maximum [Al/Fe] variation, $\Delta$max([Al/Fe])) for GCs.}
\begin{center}
\begin{tabular}{lccccc}
\hline
Cluster & [Fe/H] & $\Delta$[Al/Fe] & $\sigma$([Al/Fe]) & $\Delta$max([Al/Fe]) & Reference\\
& (dex) & (dex) & (dex) & (dex) & \\
\hline
47 Tuc  (NGC 104) &  -0.77 &  0.27 &  0.16 &  0.67 & (4)\\
NGC 288  &  -1.30     &  0.12 &  0.08 &  0.28 & (4)\\
NGC 362  &   -1.17    &  0.28  &  0.18 &  0.56 & (8)\\
NGC 1851 &  -1.18 &  0.31 &  0.20 &  0.59 & (6)\\
M~79 (NGC 1904) &  -1.57 &  0.09  &  0.07 &  0.20 & (4)\\
NGC 2808 &  -1.15 &  0.71 &  0.46 &  1.32 & (9)\\
NGC 3201 &  -1.53     &  0.57 &  0.29 &  0.81 & (4)\\
NGC 4833 &  -2.01 &  0.57 &  0.32 &  0.81 & (11) \\
M 3 (NGC 5272) & -1.54 & 0.76 & 0.33 & 1.26 & (15)\\
M~5 (NGC 5904) &  -1.34 &  0.48 &  0.28 &  0.82 & (4)\\
M~80 (NGC 6093) &  -1.79     &  0.57 &  0.33 &  1.22 & (12) \\
M~4 (NGC 6121) &  -1.17 &  0.10 &  0.05 &  0.20 & (4)\\
NGC 6139 &  -1.59 &  0.69  &  0.25 &  1.17 & (16) \\
M 107 (NGC 6171) &  -1.03      &  0.11  &  0.07  &  0.20 & (4)\\
M 13 (NGC 6205) & -1.57 & 0.77 & 0.36 & 1.16 & (15)\\
M 12 (NGC 6218) &  -1.33 &  0.35 &  0.17 &  0.66 & (4)\\
M 10 (NGC 6254) &  -1.57 &  0.41 &  0.22 &  0.60 & (4) \\
NGC 6388 &  -0.44       &  0.51 &  0.23 &  0.75 & (2)\\
NGC 6441 &  -0.39       &  0.15  &  0.13 &  0.37 & (13,14) \\
M~54 (NGC 6715) &  -1.51 &  0.95 &  0.57 &  1.37 & (5)\\
NGC 6752 &  -1.51    &  1.06  &  0.43 &  1.65 & (1,7)\\
M 55 (NGC 6809) &  -1.93    &  0.27  &  0.17 &  0.52 & (4)\\
M 71 (NGC 6838) &  -0.84 &  0.29  &  0.14 &  0.50  & (4) \\
M 15 (NGC 7078) & -2.36 & 0.57 & 0.30 & 0.93 & (15)\\
Terzan 8  &      -2.27 &  0.23 &  0.17 &  0.50 & (10)\\
\hline

\end{tabular}
\end{center}
\label{tab1}
(1) \cite{Carretta07}; (2) \cite{Carretta07-6388}; (3) \cite{Carretta09a}; (4) \cite{Carretta09}; (5) \cite{Carretta10b}; (6) \cite{Carretta11}; (7) \cite{Carretta12}; (8) \cite{Carretta13}; (9) \cite{Carretta14-2808}; (10) \cite{Carretta14-Ter8}; (11) \cite{Carretta14-4833}; (12) \cite{Carretta15}; (13) \cite{Gratton06}; (14) \cite{Gratton07}; (15) \cite{Meszaros15}; (16) \cite{Bragaglia15}
\end{table*}%

In Fig. \ref{Al:models} we show the model spectra of RSG with solar Al abundance, i.e. [Al/Fe] = 0.0 dex; and Al abundances enhanced by 0.1, 0.3 and 0.7 dex, corresponding to the synthetic spectrum expected for young GCs with moderate, intermediate and extreme [Al/Fe] ranges respectively.  The $\Delta$[Al/Fe] expected for each young GC is also found in the figure.

We conclude that if in NGC 1705: 1 there were RSGs with an extreme [Al/Fe] spread, i.e. spreads similar to the expected for GCs like NGC 2808 and 6752 at young ages, we would expect to see differences between the RSGs and NGC 1705: 1 in Fig. \ref{Al:RSG}, similar to those found between solar-scaled ([Al/Fe] = 0.0 dex) and extreme ([Al/Fe] = 0.7 dex) RSG models from Fig. \ref{Al:models}. We can exclude such extreme spreads in [Al/Fe] at high confidence, as the Al lines appear to be consistent with the SMC stars to within $\pm0.3$ dex.

\subsection{[Al/Fe] spreads: results from YMCs in the context of GCs}
 
In this section we compare the observed [Al/Fe] spreads observed in GCs with our constraints on the maximum [Al/Fe] spread consistent with our observations of NGC 1705: 1.
 
 

\begin{figure}
\includegraphics[width= 84mm]{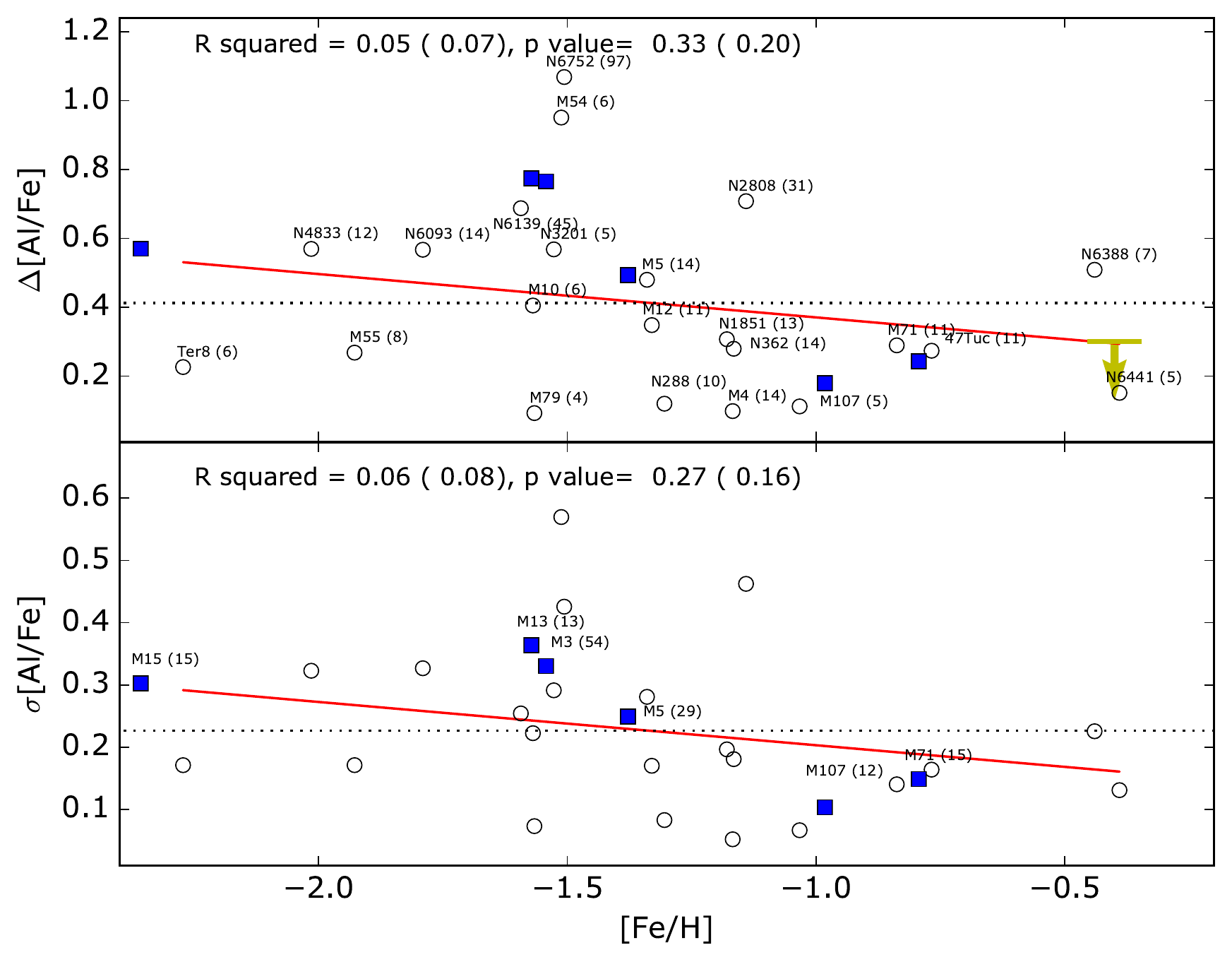}
\caption{Top panel: $\Delta$[Al/Fe] as a function of [Fe/H]. The upper limit for the Al spread in NGC 1705: 1 according to our analysis is shown as an upper limit. Bottom panel: Standard deviation $\sigma$ of the [Al/Fe] values for the stars of these clusters. The open circles represent the abundances derived by Carretta and collaborators, while the blue squares represent the Meszaros et al. APOGEE sample (blue squares). In brackets we show the number of stars with available [Al/Fe]. The red line is a linear fit to the Carretta data. The Pearson's correlation coefficient is show on the top left corners of both panels, in both cases the (anti)correlation is very weak. We also show the p-value between these distributions ([Fe/H] vs. Al-spread) and one with no correlation (i.e., Pearson's correlation coefficient = 0). We conclude that the observed Al-spread vs. metallicity distribution does not show a significant difference with respect an uncorrelated distribution, i.e. p-values>0.25. The values in parentheses are computed with the inclusion of the GCs M~3, M~13, and M~15 to the Carretta sample. Note that M~5, M~71 and M~107 are in common between the two datasets, thus we considered the $\Delta$[Al/Fe] drawn from the Carretta sample to perform the fit.}
\label{AlvFe}
\end{figure}

In Fig. \ref{AlvFe}, we plot the observed spread of [Al/Fe] in GCs as a function of the cluster [Fe/H]. We find \emph{only} a slight correlation between $\Delta$[Al/Fe] and [Fe/H] (this is also the case for the standard deviation of the [Al/Fe] - bottom panel cf. Fig. \ref{AlvFe} caption). We have also overplotted as a yellow upper limit our results for NGC 1705: 1. It seems that this YMCs is in agreement with what is expected for (old) GCs of similar metallicity.
 
\section{Discussions and conclusions}
\label{sec:conc}

From our differential analysis of the integrated \jb\ spectrum of NGC 1705: 1 presented in \S\ref{sec:GCmod}, we are not able to distinguish if this YMC is more consistent with a solar [Al/Fe] content or if it has a moderate or intermediate [Al/Fe] enhancement range in their RSGs (i.e. [Al/Fe] = 0.1, 0.3 dex; respectively), like the one observed in GCs of similar metallicity cf. Fig. \ref{AlvFe}. In principle we can not exclude any of these possibilities (cf. Fig. \ref{Al:models}). We note that from the homogeneous sample presented in Table \ref{tab1}, we find that 11 clusters ($\sim44\%$) namely (NGC 2808, NGC 3201, NGC 4833, M 3, M 80, NGC 6139, M 13, NGC 6388, NGC 6752, M 54, NGC 7078) have extreme [Al/Fe] spreads, i.e. $\Delta$[Al/Fe] > 0.5 dex, while the others have intermediate or moderate [Al/Fe] spreads.
 So according to this sample there is a 44:56 chance that the RSGs producing most of the NIR light in this cluster have an intermediate or moderate spread in [Al/Fe]. Going through a similar analysis on a larger sample of YMCs will allow us to find YMCs with extreme [Al/Fe] spreads (like GCs NGC 2808 or NGC 6752), if such objects were to exist.

Alternatively if the analysis of a robust sample of YMCs were to yield the same result, i.e. all clusters having solar 
 [Al/Fe] abundances, this would lead to the following possibilities:

\begin{itemize}
\item All YMCs do host MPs, but only display moderate [Al/Fe] spreads in their RSGs. This is highly unlikely as GCs show different levels (moderate, intermediate and extreme) of [Al/Fe] spreads.

\item GC stars in the RSG evolutionary stage do not show MPs. 
However, there is no reason in principle for this to happen, as there is evidence of MPs in GC stars at all evolutionary phases cf. P15.

\item The mechanism producing extreme [Al/Fe] spreads only kicks in (or makes them evident) after $\sim15$ Myr, i.e. the age of this cluster. As is the case for the \cite{D08} scenario to explain MPs in GCs (happening after few tens of Myr).

Or alternatively, for some reason the MPs do not become visible until after $\sim10$ Gyr of evolution. As it has not been found in open or intermediate age clusters so far (cf. \S\ref{sec:intro}).


\item The enriched population is only found in the $\sim$0.2--0.8 \msun\ stars of massive/dense clusters, i.e. only the stars alive and to which we have access in GCs. This idea has been suggested by some scenarios trying to explain the origin of MPs, (e.g. \citealt{D08,Bastian:2013p2152}), however these scenarios do not seem to work cf. \S\ref{sec:intro}. In spite of that, it is still a viable option and could be readily tested in the faintest MS stars of younger massive/dense clusters like intermediate-age clusters. A first approach to test this scenario is to see if it is possible to detect, beyond observational uncertainties, a splitting/broadening in the lower MS of young (massive/dense) clusters caused by an enriched population of stars. For this to be possible, the photometric observations of young massive/dense clusters must be carried on with the appropriate set of filters (cf. \citealt{Sbordone11}; P15). An alternative would be direct spectroscopic observations of these faint stars, however these observations are not possible with the current instruments. Nevertheless, if an enriched population is eventually found among the $\sim$0.2--0.8 \msun\ stars of young massive/dense clusters, one would be left with an additional problem, that is, to explain why there is a threshold in the mass for this phenomenon.

\item Finally, there is the possibility that none of the stars (even the low mass ones) of open clusters, YMCs and intermediate-age clusters show GC-like enrichment, and GC stars are indeed special. Then, these anomalies might be due to some special condition/mechanism (as yet unknown) only found in the early universe, $z>2$, where/when GC formed. This condition/mechanism, if it exists, has been overlooked so far in all GC formation scenarios to date. However, we know that it should only affect stars in massive/dense systems like GCs, and not be just a parameter of time, as only $\sim3\%$ of the field stars in the halo (coeval with GCs) show such abundance patterns, and these stars are thought to be GC escapees cf. \cite{Carretta10a,Martell10,Martell11,Ramirez12,Lind15}. 
But at the same time, it should be avoiding to act (or an additional mechanism should prevent it to do so) in systems somewhat more massive than GCs, like dwarf galaxies, where no evidence of such enrichment has been found (cf. \citealt{Tolstoy09,Carretta10}).
\end{itemize}

The study of a broader sample of clusters could set once and for all long standing questions like: are YMCs  and GCs objects of the same nature? And do both share these peculiar abundance patterns? If it turns out to be that both are the same kind of objects, only observed at different ages, this would suggest a common evolution of massive/dense clusters. This would represent a huge step forward in the understanding of the formation of clusters in the early universe, as we would have a more accessible way to get data (from YMCs in nearby galaxies) to carry out studies, compared to the challenging observations of the high redshift universe. On the other hand, if YMCs prove to have none of these abundance patterns, analysing the difference between them could lead us to reveal this unknown condition/mechanism that could be responsible for the MPs.

Additionally, we note that there have been two studies on YMCs where a detailed abundance analysis of the integrated $H$- and $K$-band spectra yielded an [Al/Fe] consistent with moderate/intermediate spreads ($\Delta\mbox{[Al/Fe]}\ge0.5$ dex). \cite{Larsen06} analyzed NGC 6946-1447, a young ($\sim10-15$ Myr), massive ($\sim 1.7\times 10^6$ \msun) cluster in the near by spiral galaxy NGC 6949, and found an abundance of [Al/Fe]$=0.25\pm0.18$ dex. While \cite{Larsen08} found [Al/Fe]$=0.23\pm0.11$ dex for NGC 1569-B, a young ($15-25$ Myr) massive ($4.4\times10^5$ \msun) cluster in the dwarf irregular galaxy NGC 1569. Both studies are consistent with the results of our differential analysis, i.e. no evidence of extreme Al spreads in YMCs.

We note that if this unknown condition/mechanism is to be found, we should see how it would affect the constraints placed by the studies of YMCs on the scenarios proposed for the origin of MPs in GCs.

Finally, on a cautionary note, we stress that part of our analysis is based on two assumptions: 1) that young GCs had the same [Al/Fe] distribution, as observed today, and 2) all RSGs that dominate the NIR of the young GCs have the same brightness. These assumptions need not necessarily be correct. For instance if the ratio of enriched to non-enriched stars changes significantly over $\sim10$ Gyr, in such way that the non-enriched stars are strongly predominant, we might not be sensitive to detect the signatures of some few Al-enriched RSGs. The same is true if for some reason, the enriched RSGs would be fainter than the non-enriched.

\section{Acknowledgements}

NB is partially funded by a Royal Society University Research Fellowship and an European Research Council Consolidator Grant (Multi-Pop - 646928).
\bibliographystyle{mnras}
\bibliography{cz16}

\bsp	
\label{lastpage}
\end{document}